\documentclass[global,twocolumn]{svjour}
\journalname{Applied physics B}

\newcommand{\ybgion}{\textsuperscript{172}Yb\textsuperscript{+}}

\newcommand{\SoneHalf}{$S_{1/2}$}
\newcommand{\PoneHalf}{$P_{1/2}$}
\newcommand{\DthreeHalf}{$D_{3/2}$}
\newcommand{\DfiveHalf}{$D_{5/2}$}
\newcommand{\FsevenHalf}{$F_{7/2}$}

\usepackage{times}

\usepackage{textcomp}
\usepackage[tight]{units}
\usepackage{graphicx}
\usepackage{subfigure}
\usepackage{color}

\begin{document}
\sloppy

\title{A planar ion trap chip with integrated structures for an adjustable magnetic field gradient}

\author{P.~J.~Kunert, D.~Georgen, L.~Bogunia, M.~T.~Baig, M.~A.~Baggash$^{\dagger}$, M.~Johanning, and
Ch.~Wunderlich}

\institute{Naturwissenschaftlich-Technische Fakult\"at, Department Physik, Universit\"at Siegen, 57068 Siegen, Germany   \\
            \email{wunderlich@physik.uni-siegen.de}}

\titlerunning{A planar ion trap chip with integrated structures for an adjustable magnetic field gradient}
\authorrunning{P.~J.~Kunert et. al.}

\date{Received: date / Revised version: date}
\maketitle

\begin{abstract}
We present the design, fabrication, and characterization of a
segmented surface ion trap with integrated current carrying
structures. The latter produce a spatially varying magnetic field
necessary for magnetic gradient induced coupling between ionic
effective spins. We demonstrate trapping of strings of $^172$Yb$^+$ ions,
characterize the performance of the trap and map magnetic fields by
radio frequency-optical double resonance spectroscopy. In addition,
we apply and characterize the magnetic gradient and demonstrate
individual addressing in a string of three ions using RF radiation.
\end{abstract}

\maketitle

\section{Introduction} \label{sec:Introduction}

Cold trapped ions have been established as a benchmark system in
quantum information science and were used to show a variety of first
proof of principle demonstrations
\cite{NielsenChuang2000,BlattWineland2008} in this field. When
scaling up the number of qubits or ions, a widely favored solution
to limit detrimental effects of decoherence is to split the entire
quantum register into partitions of manageable size by using
segmented traps featuring loading and processor zones
\cite{Wineland1998,Kielpinski2002}, and ion transfer between zones
can be achieved in a fast diabatic manner optimized to reduce
heating \cite{Walther2012,Bowler2012}. When increasing the
complexity of such traps, planar designs are often favored, as they
benefit from elaborated micro-system fabrication techniques. These
allow for very flexible designs \cite{Seidelin2006,Pearson2006} (for
a recent review see \cite{Hughes2011}) relevant  for universal
quantum computing,  but also in the context of quantum simulations,
where customized electrode shapes can be used to realize various
lattice structures and interaction types between trapped ions
\cite{Wesenberg2008,Schmied2009,Siverns2012a}.

Ionic qubits can be manipulated with high fidelity using laser-based
gates, whereas qubits encoded into hyperfine states can also be
manipulated directly by radio frequency (RF) fields.

    \begin{figure}
        \centering
        \includegraphics[width=0.99\columnwidth]{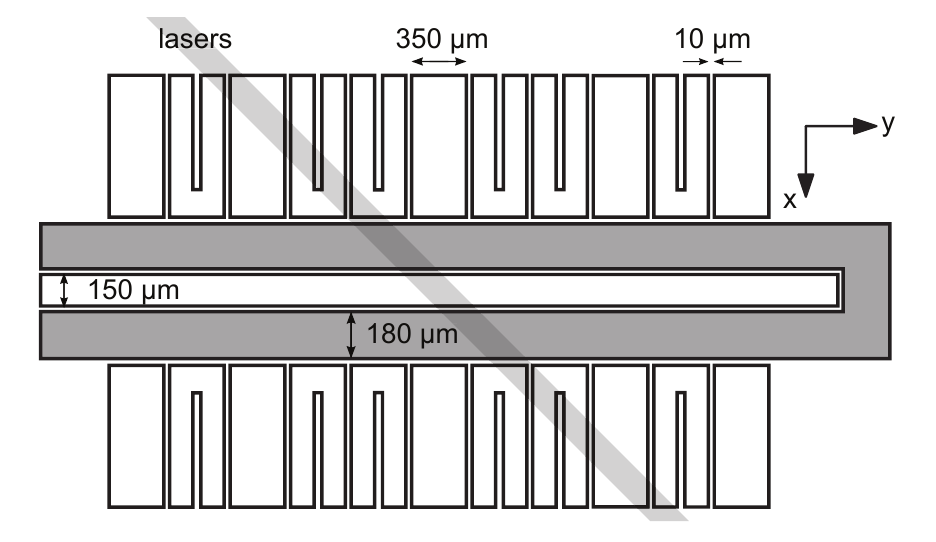}
        \caption{Schematic of the five-electrode planar trap
                geometry with integrated current loops for a flexible axial magnetic gradient
                shape. The RF trapping field is applied to the electrodes shown in grey and
                provides radial trapping. The segmented dc electrodes allow for
                tailoring of the axial electric potential for ion transport, and, due to
                slots, for application of currents for a flexible axial magnetic
                field and gradient. The diagonal grey line indicates the direction of laser beams
                used for Doppler cooling and detection of trapped ions. }
        \label{chipScheme}
    \end{figure}

One way to maintain the addressability of single ions despite their
separation being orders of magnitude below the diffraction limit is
the application of a static magnetic field gradient and exploiting
an inhomogeneous Zeeman effect
\cite{Mintert2001,Wunderlich2002,Johanning2009b,Khromova2012,Piltz2013}
which allows addressing in frequency space. In this way, low
crosstalk can be achieved \cite{Khromova2012}. For addressing of
individual ions it has also been proposed \cite{Staanum2002} and
demonstrated \cite{Navon2012} to use inhomogeneous laser fields, and
addressing has been demonstrated using oscillating microwave
gradients \cite{Warring2013}.

Coupling between internal and motional states of trapped ions --
needed for conditional quantum dynamics with several ions -- is
negligible in usual ion traps when RF radiation is applied. In the
presence of a static \cite{Mintert2001,Johanning2009b} or
oscillating \cite{Ospelkaus2011} magnetic field gradient, however,
such coupling is induced. Also, coupling between spin states of
different ions \cite{Wunderlich2002,Khromova2012,Piltz2013} arises
in a spatially varying magnetic field and is thus termed magnetic
gradient induced coupling (MAGIC).

A static gradient can be generated by permanent magnets
\cite{Johanning2009b,Khromova2012} or by current loops that allow to
introduce a time dependence. This was implemented into 3d ion trap
designs \cite{Kaufmann2011}, discussed for planar geometries
\cite{Welzel2011}, and applied for addressing in frequency space
using a laser quadrupole transition \cite{Wang2009}. The tailoring
of the interactions between ions can be achieved by shaping the
axial electrostatic trapping potential
\cite{Wunderlich2002,HWunderlich2009,Khromova2012}, but also by
changing the shape and direction of the magnetic field gradient.

In what follows, we discuss design considerations and
fabrication details for a planar trap with integrated segmented
loops which provide a magnetic field gradient whose spatial
dependence can be tailored. We present experimental results with
trapped ytterbium ions and demonstrate for the first time the
application of a magnetic field gradient for RF addressing of ions
in a planar trap.

\section{Experimental setup} \label{sec:ExperimentalSetup}

    \subsection{Trap design and fabrication} \label{sec:TrapDesignAndFabrication}

    \begin{figure}
       \includegraphics[width=0.99\columnwidth]{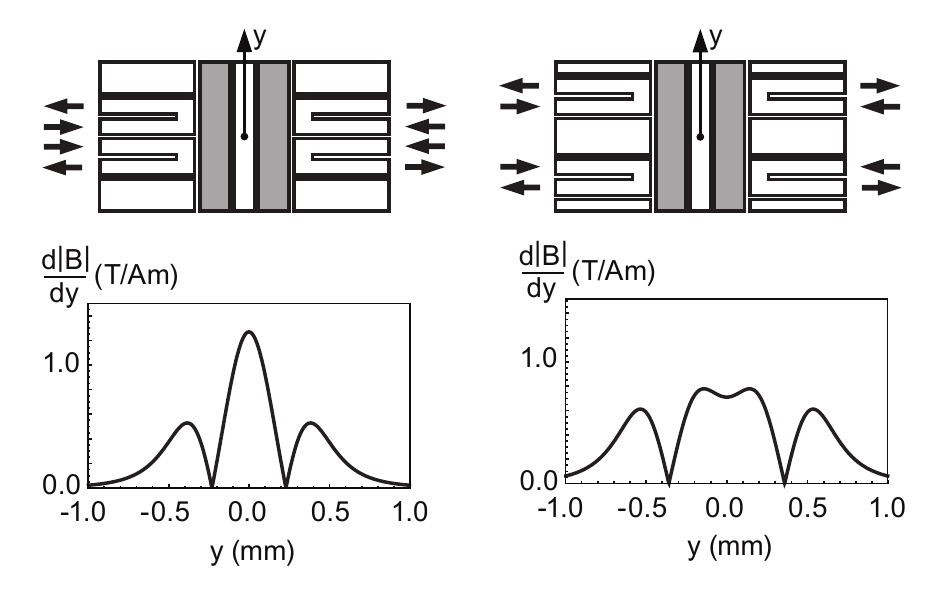}
        \caption[simAxialGrad]{Current carrying structures and magnetic field gradient of the planar ion trap.
                Upper panel: Electrode structure. RF electrodes are shown in grey (width of 120~\textmu m).
                The ground electrode and the segmented DC electrodes (width of 350~\textmu m) are shown in white.
                The DC electrodes are split allowing for the application of a current (indicated by arrows).
                Lower panel: Here, two possibilities to generate a magnetic field gradient are shown corresponding
                to the currents indicated by arrows in the upper panel. The simulated magnetic gradient (for a
                current of 1~A) is plotted as a function of the axial coordinate $y$.
                Left: quadrupole configuration, right: stretched quadrupole configuration.}
        \label{gradientsim}
    \end{figure}

    \begin{figure}
        \centering
        \includegraphics[width=0.99\columnwidth]{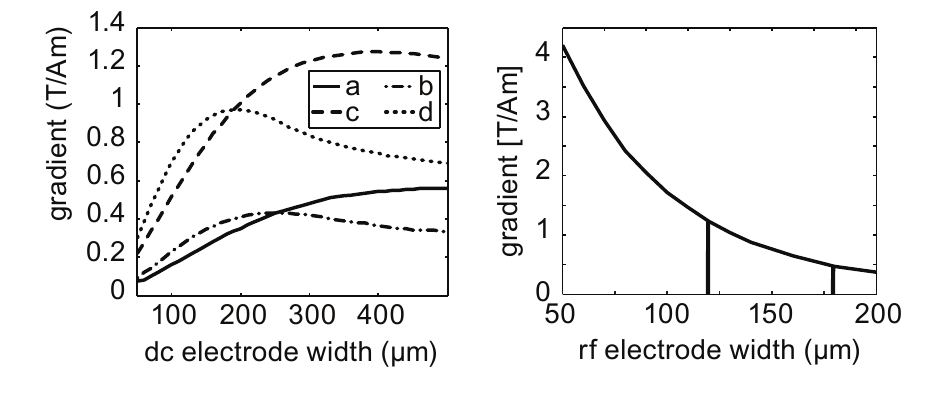}
        \caption[simGradSize]{Simulated axial magnetic field gradient for varying electrode widths.
                Left: The gradient is shown as a function of the width of the segmented dc electrode
                with fixed RF electrode width at 180~\textmu m (a, b) and 120~\textmu m (c, d),
                respectively, for two different current configurations: quadrupole (a, c) and streched
                quadrupole (b, d) as motivated in fig.~\ref{gradientsim}. For every fixed RF electrode
                width a segmented dc electrode width can be found which maximizes the gradient.
                Right: The gradient as a function of the RF electrode width (and in turn the trapping
                height) is shown; here, segmented dc electrode width is fixed at 350~\textmu m.
                By reducing the trap dimension, gradients higher than 4~T/Am are predicted. The RF
                electrode widths presented in the left part are marked by vertical lines in the right
                plot.}
        \label{grad_width}
    \end{figure}

The trap presented here is a symmetric five-electrode planar trap
design \cite{Chiaverini2005}. The outer dc electrodes are segmented
to provide axial confinement and allow for axial ion transport.
Numerical simulations based on analytical solutions for planar traps
\cite{Oliveira2001,House2008} were carried out for various electrode
dimensions to maximize the trap depth at given RF amplitude, and we
choose the electrode width as 180~\textmu m for the radio frequency
electrodes and 150~\textmu m for the middle control electrode (see
fig.~\ref{chipScheme}). Eleven dc electrode pairs allow to transport
ions over a range of several millimeters and define several
independent trapping regions. In addition, we
integrated current loops for the creation of inhomogeneous fields to
allow for MAGIC.


So far, approaches for integrated current loops for five electrode
planar traps designs used a current through a patterned center wire
\cite{Welzel2011,Wang2009} to create an inhomogeneous magnetic
field. In that case the shape and the position of the gradient are
predetermined by the design of the electrode/current loop. Here, we
introduce a new approach where several segmented dc electrodes are
slotted. Applying currents with individual magnitude and direction
through these micro-structured integrated current loops provide a
magnetic field gradient with a variable shape and strength along the
axial direction.

The design introduced here makes use of up to twelve segmented dc
electrodes to generate the magnetic field (see
fig.~\ref{chipScheme}). The width of these electrodes primarily
determines the shape and peak strength of the gradient for a given
current. Two basic current patterns are used here to optimize the
electrode/current loop geometry (compare fig.~\ref{gradientsim}): in
quadrupole configuration (fig.~\ref{gradientsim} left) the current
is applied to any neighboring current loops in a symmetric fashion,
resulting in a strong peak gradient strength for a given current.
When the same pattern is applied in stretched quadrupole
configuration, the gradient extends over a larger region, at the
expense of a lower peak gradient strength (see
fig.~\ref{gradientsim} right). The electrode width affects the peak
gradient value in both scenarios and numerical simulations using
Biot-Savart's law were used to find the electrode width of
350~\textmu m to be a good a compromise between the peak strengths
for both patterns (see fig.~\ref{grad_width} left). The gradient
depends also on the width of the RF electrodes, as illustrated in
fig. \ref{grad_width} showing the peak gradient strength in
quadrupole configuration for a dc electrode width of 350~\textmu m.
A flexible shaping of the gradient can be achieved as any current
pattern with varying current strength can be applied, resulting in
the weighted sum of the individual gradients.

    \begin{figure}
        \centering
        \includegraphics[width=0.99\columnwidth]{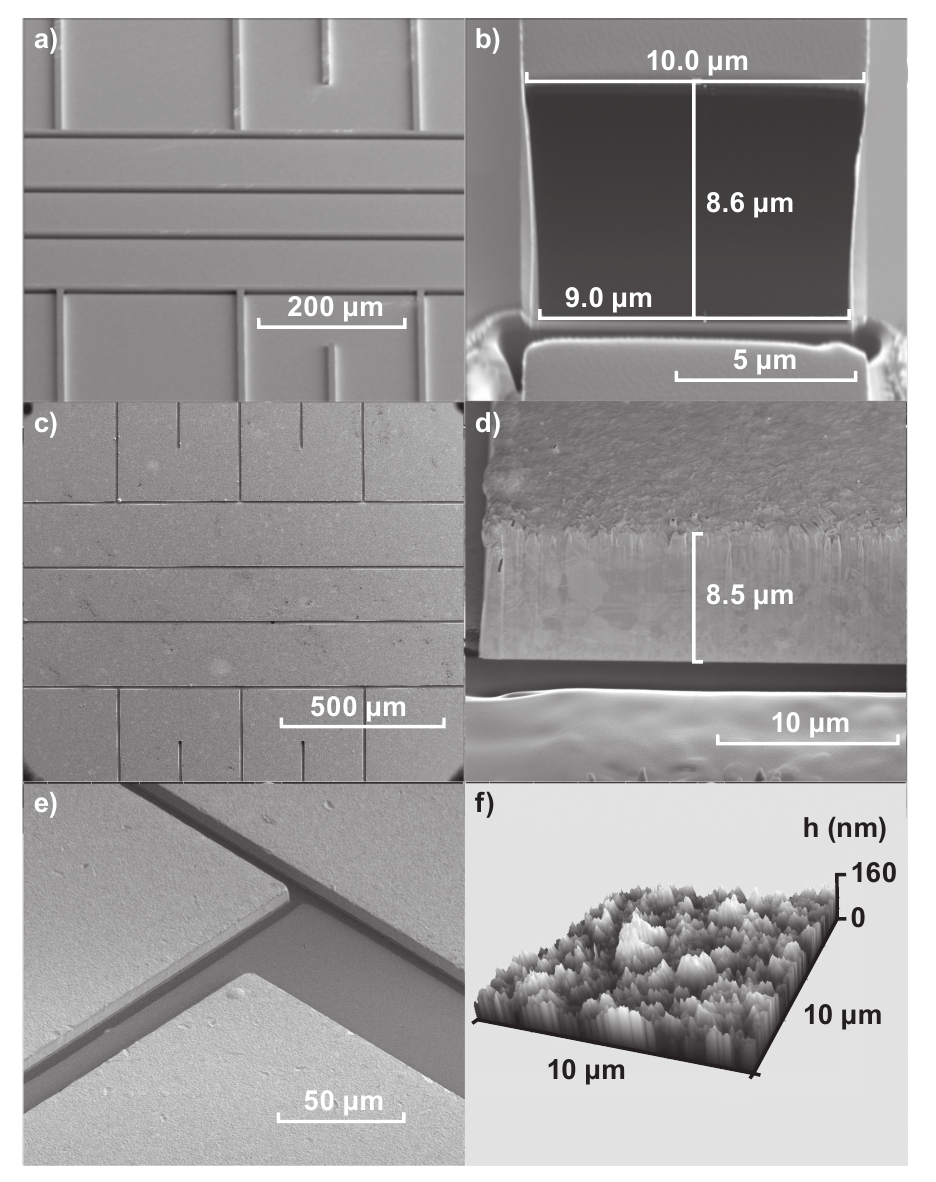}
        \caption[chipProd]{Focussed-ion-beam imaging of the relevant production steps;
                a) Resist structure after photolithography that defines the gaps between the electrodes in the following electroplating step.
                In this image, the dark lines indicate resist structures that are elevated above the surface of the substrate.
                b) To determine the quality of the resist structure, we removed a part of the resist with an ion-beam and visualized the structure
                under 52$^\circ$ relative to the substrate surface. In this way the resist thickness (8.6~\textmu m) and the widths at the top
                (10~\textmu m) and bottom (9~\textmu m) can be determined.
                c) Electroplated electrodes.
                d) Cut through one electroplated electrode and measured gold height to 8.5~\textmu m.
                e) Structure after physical etching with a still existing chrome layer (dark grey) between the gold electrodes (light grey).
                f) A surface roughness of 20~-~30~nm rms is measured with an AFM for different
                chips and different positions on the chips. The figure shows one sample for illustration.}
        \label{chipprod}
    \end{figure}

The materials chosen for the trap chip were selected for their
compatibility with ultra-high-vacuum, high RF voltages and high
currents up to several Ampere to obtain large trap depths and
magnetic field gradients, low RF losses and high thermal
conductivity to remove heat intake efficiently. The chip substrate
is made of sapphire, as this material allows for high RF amplitudes
due to a high electrical resistance and low absorption of RF power.
Furthermore, a good surface roughness of around 3~nm can be
obtained. The adhesion of the electrode material (in our case gold)
is essential and we improve it by an additional intermediate
adhesion layer of chromium. Any thermal load, either by RF losses or
from high currents required for large magnetic gradients, is
efficiently dissipated due to the large thermal conductivity
(45~W/mK) of sapphire.

The largest possible magnetic field gradient is given by the damage
threshold of the integrated coils due to ohmic heating, so we aim to
obtain a high damage threshold by having a low resistivity (using
thick electrodes made of gold, which has a low specific resistivity)
and by quickly removing the heat by the good thermal conductivity of
the gold electrode and the sapphire substrate.

We create the trap electrodes by sputtering a 10~nm chrome adhesion
layer followed by a 50~nm gold seed layer. Before every sputtering
process a physical etching step cleans and smoothes the processed
surface. The layer thickness obtained by standard sputtering or
evaporating processes is usually limited to around 1~\textmu m.
Thicker structures can be obtained by electroplating. The gaps
between the electrodes are defined by optical lithography
(fig.~\ref{chipprod} a,b). In this process the wafer is spin coated
with negative photo resist (AZ15nXT) with a height of 8.5~\textmu m.
A baking step reduces the solvent before the resist is covered with
a photolithography mask and exposed with uv light (i-line of
Hg~lamp). Another baking step crosslinks the resist. Then, after a
chemical developer (AZ826) has removed the exposed resist, a
cleaning process with oxygen plasma removes unintended residual
resist. The resist structure obtained in this process yields nearly
vertical edges and high aspect ratios (fig.~\ref{chipprod} a,b). To
determine the quality of the resist structure, we removed a part of
the resist with an ion-beam and visualized the structure under
52$^{\circ}$ relative to the surface. In this way the resist
thickness and the widths at the top and the bottom can be
determined. It can be seen in fig.~\ref{chipprod} b) that these
widths differ only by roughly 10~\% (9~\textmu m versus 10~\textmu
m).

Electroplating is carried out using an open bath (Metakem SF6) under
atmospheric conditions. The bath is temperature stabilized and
pH-value controlled and can operate with current densities as low as
the minimum specified value for the solution
(1~mA/cm\textsuperscript{2}). At this current density, we obtain a
gold deposition rate around 60~nm/min and a smooth surface quality
with an rms roughness around 25~nm (see fig.~\ref{chipprod} f). We
electroplate gold layers up to a thickness of 8.5~\textmu m (see
fig.~\ref{chipprod} c,d). The resist is removed after electroplating
using wet etching (DMSO) before the seed layers can be physically
etched with an argon plasma (fig.~\ref{chipprod} e) which can be
controlled on a nanometer scale.

    \begin{figure}
        \centering
        \includegraphics[width=0.99\columnwidth]{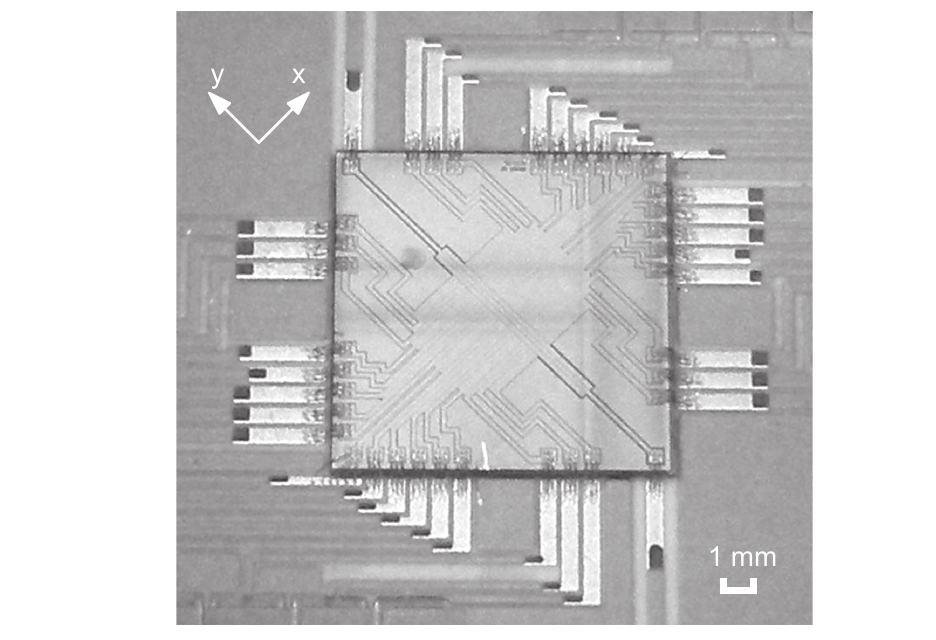}
        \caption[fig5]{Assembled electroplated ion trap chip with an edge length
                        of 11~mm onto an alumina carrier with printed silver-palladium wires.
                        Every electrode is ball bonded six times with 50~\textmu m gold wires
                        to gold coated bond pads on the carrier.}
        \label{chip_bw.png}
    \end{figure}

The trap is mounted on a custom made chip carrier made of
alumina for its high thermal conductivity of 25 W/mK and its
machinability with pulsed CO$_{2}$ or Nd:YAG lasers. We use thick
film technology \cite{Gupta2005} to print wires, resistors and
capacitors onto the chip carrier to integrate low pass filters for
each dc electrode with an cut-off frequency in the kHz range.
Similar chip carriers have been demonstrated before and can also be
used as a vacuum interface \cite{Kaufmann2011}. The maximum current
is at present limited by the resistance of the feed wires on the
carrier which is near 8~$\Omega$ for a single loop.

The trap depth can be increased by mounting a conductive mesh at a
distance of a few millimeters parallel to the trap surface and applying a
positive voltage \cite{Brown2007}. Such an electrode also reduces
the effect of stray charges of the optical viewports used for the
detection of the ion (see sec.~\ref{sec:LaserSystem}). Here, we use
instead a glass slide made of borosilicate glass with a thickness of
60~\textmu m and coat it with 100~nm layer of transparent, but
conductive indium-tin-oxide (ITO) \cite{Yan2009} by sputtering. In
this way, the glass slide can be connected to a voltage supply and
can be used as a transparent electrode (transmission 70~\% at
369~nm).

    \subsection{Laser system and detection} \label{sec:LaserSystem}

The laser system is, apart from minor modifications, as it has been
used to trap ions in a 3d segmented linear trap with a built-in
magnetic gradient coil and is described in \cite{Kaufmann2011}. All
lasers are external cavity diode lasers, locked to temperature and
pressure stabilized low drift medium finesse Fabry-Perot cavities
(with finesses in the range of 50\dots 200). The lasers are fibre
coupled and overlapped using dichroic mirrors before they enter the
vacuum chamber. All wavelengths are simultaneously determined using
a home-built scanning Michelson interferometer which allows for a
relative accuracy of $\delta \lambda$/$\lambda$~$\approx 10^{-8}$
corresponding to a few tens of MHz for all our lasers. Using this
lambdameter alone, one can set the wavelengths precisely enough to
see ionic fluorescence. A beam of neutral atoms is generated by
ohmic heating of a miniaturized atomic oven. The atoms are
photoionized using  two-step photo-ionization with a resonant first
step which is driven using a laser near 398~nm
\cite{Balzer2006,Johanning2011,Onoda2011}. From there, the cooling laser (see
below) near 369~nm drives the transition into the ionization
continuum. The laser beams are aligned parallel to the trap surface
and are adjusted under 45$^\circ$ relative to the trap axis to
achieve Doppler cooling of radial and axial modes. The out-of-plane
motion is not or only weakly cooled due to fringe potentials.

    \begin{figure}[t]
        \centering
        \includegraphics[width=0.99\columnwidth]{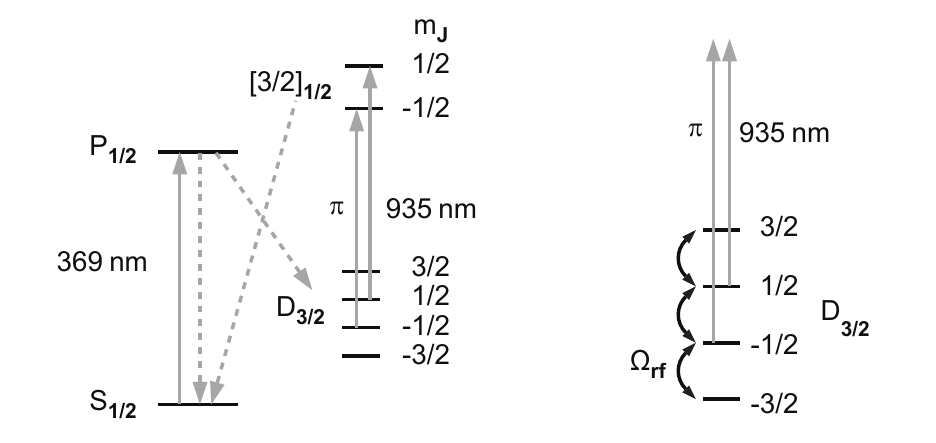}
        \caption[fig6]{Relevant energy levels of \ybgion (not to scale).
                Left: The electric dipole transition between the S$_{1/2}$ ground state
                and the P$_{1/2}$ excited state near 369~nm is used for
                Doppler cooling and state selective detection by detecting
                resonance fluorescence with a photomultiplier or an intensified
                CCD camera (termed "cooling fluorescence"). Laser light near 935~nm coupling
                the metastable state D$_{3/2}$-state to the $[3/2]_{1/2}$ state allows
                for control of optical pumping into the D$_{3/2}$-state.
                Right: RF radiation ($\Omega_{\rm{rf}}$) couples the states populated by
                optical pumping to those which are depopulated by the repumping laser near
                935~nm (see sec.~\ref{sec:RfOpticalDoubleResonanceSpectroscopy}).}
        \label{fig:levels}
    \end{figure}

The relevant energy levels of \ybgion are shown in
fig.~\ref{fig:levels}. For cooling and state-detection, we use the
resonance transition between the \SoneHalf- and the \PoneHalf-state
near 369~nm. Spontaneous decay into the \DthreeHalf-state requires
an additional laser near 935~nm for repumping into the ground state
\SoneHalf. Collisions with background gas with sufficient energy can
mix the \DthreeHalf-state with the \DfiveHalf-state. This state can
decay into the \FsevenHalf-state which has been used in clock
experiments and has a predicted unperturbed lifetime of several
years \cite{Blythe2009}. Considering the background pressure in our
experiments $<3\cdot 10^{-11}$~mbar, this collision-assisted
loss rate is in the range of sub-milli Hertz and thus we do not
repump this state with an additional laser but in such cases drop
the ion from the trap and reload.

Fluorescence from trapped ions is collected with a large numerical
aperture lens-system (NA=0.4) which is optimized for diffraction
limited imaging of ions over a large field of view \cite{Schneider}.
A schematic cross section of the light gathering system  can be
found in \cite{Kaufmann2011}. The fluorescence is discriminated
against stray-light from the trap chip by a telecentric imaging
system. This setup located in an aluminum box anodized for high
absorption ($\approx 90\%$ absorption for 369~nm laser light) includes three
planes where high absorption coated moveable razor blades ($95\%$ absorption
for 369~nm laser light) are mounted. Two blade pairs form a
rectangular aperture localized in the focal plane of the imaging
objective. Ions are imaged via an extension lens onto an EMCCD
camera (Andor iXon\textsuperscript{+}). A third pair of blades aid
in blocking light scattered  from objects originating at different
locations near the trapping region. Thus, the signal-to-background
ratio can be improved. Stray-light from all lasers with wavelengths
different from 369~nm is effectively suppressed using a narrow
band-pass filter with a spectral width of 6~nm (FWHM) in front of the camera.

    \subsection{Electrical signals} \label{sec:ElectricalSignals}

The RF voltage required to trap ions is generated by a signal
generator which is amplified and fed into a helical resonator and
the details of the setup are discussed in this section. The helical
resonator follows the general concept reported in
\cite{Macalpine1959,Vizmuller1995,Zverev1967} and is designed as an
autotransformer. This approach yields lower insertion loss, but also
results in lower Q-factors compared to the approach described in
\cite{Siverns2012}. We carefully designed the resonator for
mechanical stability and wound the helix on a threaded low loss
dielectric tube (PTFE). The mechanical stability results in a low
drift of the resonance frequency of $\pm$30~Hz over several hours.
This was measured using a capacitive load ($\approx$30~pF) which is
comparable to our trap including connectors ($\approx$35~pF). The
insertion point of the primary coil, which is critical for impedance
matching, is realized as a slider which can be firmly fixed with a
set screw with good electrical contact, but, at the same time, can
be moved with little effort. The tube can be filled, also partially,
by a dielectric to tune the resonance frequency of the circuit. The
frequency tuning range is found to be on a percent scale of the
resonance frequency (for up to 80~\% filling with PTFE) and can
alternatively be achieved by varying the load, for instance, by a different
length of the cable connecting the resonator to the trap.

Even with a Q-factor near 80 we find that, due to low insertion
losses, the resonator can drive a trap like the one described in
\cite{Kaufmann2011} with a RF power of only 0.25~W, which is readily
delivered by the signal generator (as, for example, a Rigol DG1012)
and requires no further amplification. In the experiments presented
here, the RF amplitude is generated by a frequency generator (Hameg
HM8032) and amplified with a Kalmus~110C by 40~dB to a power of
approximately 0.4~W. The helical resonator boosts the
peak-to-peak-voltage at the trap frequency of 14.7~MHz. The voltage
is fed into the trap and simultaneously monitored via a 1~pF
capacitance probe. The system is optimized to avoid ground loops.

A system of DAC cards (Adwin Pro~II) connected to 50 $\Omega$
drivers delivers 10 tunable voltages in the range of $\pm 10$~V.
Via jumpering (compare \cite{Kaufmann2011}) up to 75 potentials can
be routed to the trap electrodes via sub-d connectors.

\section{Trap characteristics} \label{sec:TrapCharacteristics}

    \begin{figure}
        \centering
        \includegraphics[width=0.99\columnwidth]{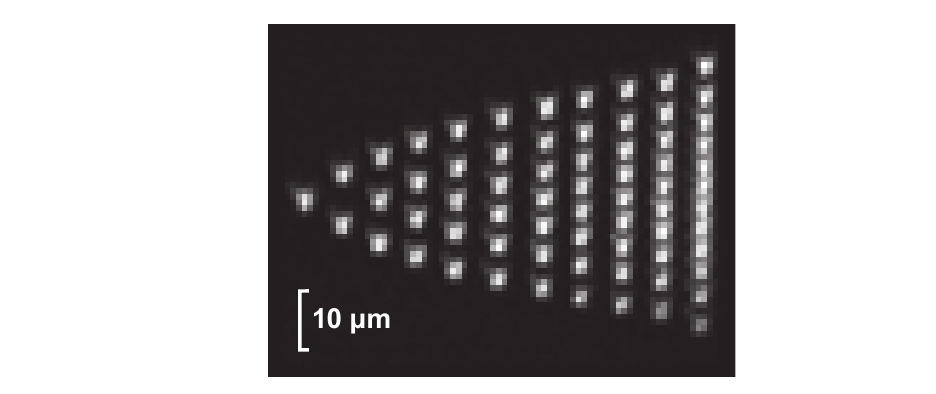}
        \caption[ionChains]{A composite picture of trapped $^{172}$Yb ion chains.
                        Between one (leftmost columns) and twelve (rightmost column) ions are trapped.
                        Each of the twelve individual ion chain pictures is taken from one single loading procedure.}
        \label{ionchain}
    \end{figure}

We demonstrate trapping $^{172}$Yb$^{+}$ ions in our planar trap
(fig.~\ref{ionchain}). The storage time with laser cooling but
without repumping the $F_{7/2}$-state is several hours for single
ions and several 10~minutes for ion chains up to 10 ions.

A measured trapping height of (160 $\pm$ 10)~\textmu m is in
agreement with the numerical simulation of the trapping potential.
The measured ion-ion distance for two ions is 10~\textmu m (taking
into account the independently determined magnification of the
detection system). Ions are stable for RF peak-to-peak amplitudes
between 150~V$_{\rm pp}$ and 400~V$_{\rm pp}$. With a typical
trapping amplitude of 250~V$_{\rm pp}$ and a trap drive frequency of
14.7~MHz, the stability parameter \cite{Chiaverini2005}

\begin{equation}
    q = \frac {2Q V_{\rm rf}} {m \Omega^{2} r_{0}^{2}}
\end{equation}

is determined to be 0.22, with the charge $Q$, RF amplitude $V_{\rm
rf}=V_{\rm pp}/2$, trap drive frequency $\Omega$ and trap geometry
factor $r_{0}$. The trap depth \cite{Chiaverini2005}

\begin{equation}
    \Psi_0 = \frac {Q^2 V_{\rm rf}^{2}} {4m \Omega^{2} r_{0}^{2}}
\end{equation}

is determined to be 73~meV.

    \begin{figure}
        \centering
        \includegraphics[width=0.99\columnwidth]{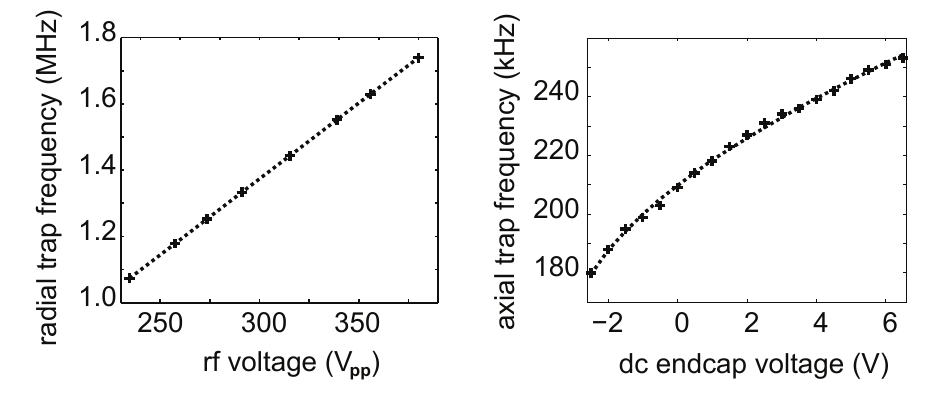}
        \caption[fig8]{Measured trap frequencies as a function of applied voltages.
                    Left: Measured radial trap frequency which is directed parallel to
                    the trap surface with 1~V$_{\rm dc}$ applied to the endcap electrodes.
                    Right: Measured axial trap frequency at 250~V$_{\rm pp}$ RF amplitude.
                    The errors for measured frequencies are 0.5~kHz and for applied
                    voltages 0.01~V$_{\rm dc}$ and 1~V$_{\rm pp}$ respectively.}
        \label{trapfreq}
    \end{figure}

We measure trap frequencies by resonant heating, which occurs,
when the trap frequency coincides with the frequency of a sinusoidal
'tickling' signal applied to one dc electrode. The motional
frequencies are determined for the axial direction in the range from
180~kHz to 250~kHz (fig.~\ref{trapfreq} left) and for the radial
direction parallel to the trap surface between 1.0~MHz and 1.8~MHz
(fig.~\ref{trapfreq} right).

    \begin{figure}
        \centering
        \includegraphics[width=0.99\columnwidth]{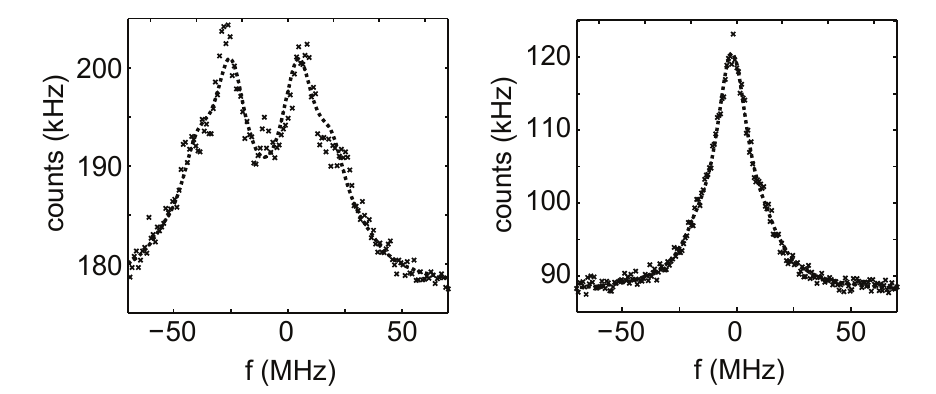}
            \caption[fig9]{Micromotion compensation analysed with
                    absorption spectra of the 935~nm repump laser (see text). Left: Before micromotion
                    compensation several motional sidebands at multiples of the trap drive frequency
                    of 14.7~MHz are visible.
                    Right: Micromotion sidebands are suppressed in the spectrum after micromotion
                    compensation was carried out and predominantly the carrier is visible.
                    The function fitting the dataset is given by the lineshape calculated for
                    an oscillating ion \cite{Wineland1979}.
                    The variations in frequency and amplitude between the two spectra shown here
                    are caused by laser frequency drifts during data taking.}
        \label{mubew}
    \end{figure}

Stray fields may prevent the ion from being trapped at the bottom of
the effective potential, where the RF electric field vanishes. In
that case the ion's driven motion results in sidebands in the
absorption spectrum that are separated from the carrier by multiples
of the RF trap drive frequency (fig.~\ref{mubew} left). To detect
and compensate this motion, the dependence of the ion fluorescence intensity
on the detuning of the 935~nm laser is analyzed. By changing potentials
applied to the segmented dc electrodes the ion can be moved slightly
along the radial direction towards the RF minimum, where the
sidebands are reduced and the carrier dominates the absorption
spectrum (fig.~\ref{mubew} right).

Background light is reduced by a telecentric imaging setup as
described above. The signal-to-background-ratio is optimized
starting with the blades initially fully open and then closing them
until the best ratio is achieved. For small apertures, both signal,
and background depend approximately linearly on the area of the
aperture and we use the intersection of the tangent to the
fluorescence rate with the fluorescence rate which saturates for
large apertures to find a working point for the blade setting
yielding a signal-to-background-ratio of $211 \pm 9$ compared to
54~$\pm$~2 with fully open blades.

\section{RF-optical double resonance spectroscopy} \label{sec:RfOpticalDoubleResonanceSpectroscopy}

We demonstrate one of the two basic effects of an inhomogeneous
magnetic field, the addressing of ions in frequency space, using the
Zeeman levels of the $D_{3/2}$ manifold (see fig.~\ref{fig:levels}).
These levels have been previously used to show addressability of
ions using RF transitions in a magnetic field gradient
\cite{Johanning2009b}.

The metastable D$_{3/2}$-state (lifetime of 52.2~ms \cite{Gerz1988})
is populated by spontaneous decay from the $P_{3/2}$-state with a
branching ratio of approximately 0.5~\% \cite{Olmschenk2007}. Thus
optical pumping into this level occurs on a microsecond time-scale
into all Zeeman levels. In order to close the fluorescence and
cooling cycle, laser light near 935 nm is applied that excites the
ion to the $[3/2]_{1/2}$-state which subsequently decays to the
ground electronic state. For cooling and detection, the polarization
of the 935 nm light has to contain at least $\sigma^+$ and $\sigma^-$
components in order to prevent optical pumping into any of the
Zeeman states of the D$_{3/2}$-level. Light near 935~nm containing 
$\pi, \sigma^+,$ and $\sigma^-$ components is achieved by using a
linearly polarized light beam incident on the ion at 45$^\circ$
relative to the quantization axis. Thus, the population from all
Zeeman substates is pumped back to the ground electronic state.
However, in order to demonstrate individual addressing we want to
first prepare an ion deterministically in one of the Zeeman
substates of the $D_{3/2}$-level by optical pumping. This process is
described in what follows.

    \subsection{State initialization} \label{sec:QubitInitializationAndReadout}

When the repumper near 935 nm is linearly polarized and its electric
field is aligned parallel to the magnetic field that determines the
quantization axis, it will exclusively drive $\pi$-transitions which
do not change the Zeeman quantum number $m_J$. Thus, population
accumulates in the Zeeman states $m_J=\pm 3/2$ that  are not coupled
to the light field, since the $[3/2]_{1/2}$-state cannot be accessed
from them by $\pi$-transitions. Both levels $m_J=\pm 3/2$  are
populated with equal probability (assuming no imperfections in the
polarization state). Once the ion is optically pumped into these two
Zeeman states, fluorescence and cooling of the ion stops. This
initialization scheme requires an electric field polarization
parallel to the magnetic field, and thus a propagation direction,
indicated by the $k$-vector of the light field, perpendicular to it.

When one circular polarization component is added to the 935 nm
light,  population will be trapped in a {\em single} Zeeman state,
for instance in the $m_J = 3/2$-state for $\sigma_+$-polarization
(compare fig.~\ref{fig:levels}). Circular polarization requires a
field vector rotating around the magnetic field and thus requires a
$k$-vector parallel to this field.

    \begin{figure}[t]
        \centering
         \includegraphics[width=0.99\columnwidth]{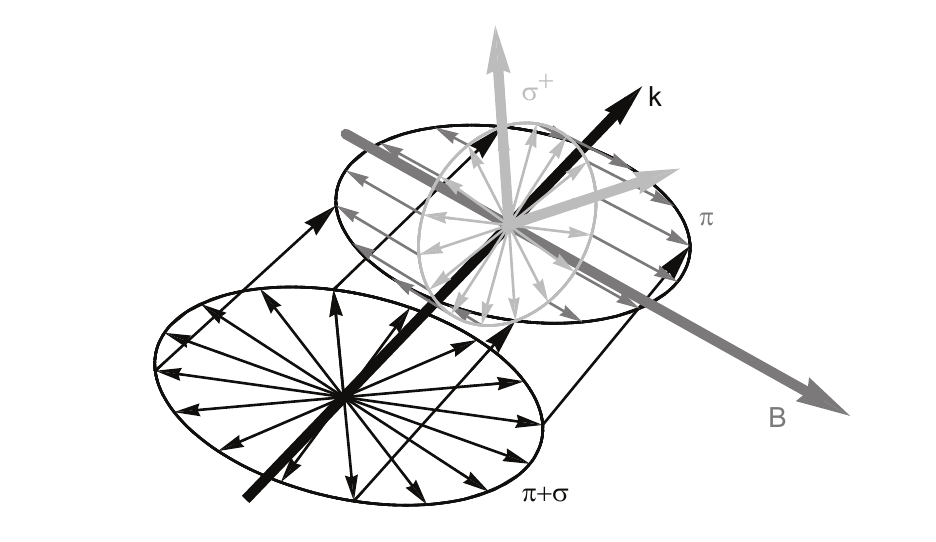}
        \caption[fig11]{Decomposition of suitable elliptical polarization into
                phase coherent linear and circular polarizations. If the projection of the elliptical polarization
                onto the plane perpendicular to the external field determining the quantization axis (B) yields
                exactly a circle, the total polarization will only drive a $\pi$
                and one $\sigma$ component.}
        \label{fig:polarization}
    \end{figure}

Instead of using two light fields for initializing the ion in one
Zeeman state, we construct the polarization state of the field for a
given $k$-vector, as shown in fig.~\ref{fig:polarization}. In this
way the qubit can be initialized in a single Zeeman state using only
one laser beam. This is a general approach, because of its
independence on the direction of the magnetic field that determines
the quantization axis.

Here, we arrange the degree of elliptical polarization such, that
the projection onto the plane normal to the magnetic field is
circular. In this way, the total light field is composed of a
linearly polarized electric field parallel to the magnetic field and
a circularly  polarized component perpendicular to the magnetic
field. Thus, population can be trapped in either one of the Zeeman
states $m_J = \pm 3/2$ (creating a ``dark'' state).


    \subsection{RF manipulation of the Zeeman states} \label{sec:RFManipulationOfThePseudoQubit}

Individual addressing of ions is demonstrated using RF-optical
double resonance spectroscopy on the D$_{3/2}$-state. The ion is
first prepared in a desired Zeeman state as described above (a dark
state). Then RF radiation is applied that brings the ion back into
the fluorescence cycle, if it's frequency is close to the resonance
that corresponds to a transition between Zeeman substates. This
resonance frequency is determined by the strength of the local
magnetic field that the ion is exposed to as explained in what
follows.

The degeneracy of its Zeeman-manifold is lifted by a magnetic field
$B=B_0 + y \;\partial_y\,B_G $ composed of an offset field $B_0$ and
an additional field with constant gradient $\partial_y\,B_G$ (the
axial direction $\vec{y}$ is the direction of the softest trap
frequency).

The magnitude of $B$ determines the resonance frequency of magnetic
dipole transitions between the Zeeman states. The (linear) Zeeman
shift $\Delta E_J$ by a magnetic field $B$ is given by $\Delta E_J =
g_J \, m_J \, \mu_B \, B$ with Land\'{e} g-factor $g_J$, magnetic
quantum number $m_J$, and the Bohr magnetron $\mu_B$. Magnetic
dipole transitions between levels with $\Delta\, m_J = \pm1$
(magnetic $\sigma$-transition) with resonance frequency

\begin{equation}
    f = \frac{g_J \, \mu_B \, B}{h} \label{eq:transition}
\end{equation}

are driven using an RF field that is generated by a dipole coil
wrapped around the light gathering optics. The magnitude of $B_0$
was chosen to be about 0.6~mT resulting in a resonance frequency $f
\approx 7$~MHz.

We generate the RF signal using a  signal generator (VFG~150). This
device allows to control the frequency, amplitude and phase and
allows for fast phase continuous and phase coherent switching
\cite{Hannemann2007,Timoney2008,Timoney2011}. The output of the
signal generator  is amplified (Kalmus~110C) with a gain of 40~dB to
a power of up to 5~Watts. The signal is applied with a loop antenna,
which forms a resonance circuit with a series capacitor. Impedance
matching is achieved by using an adjustable series resistance.

    \begin{figure}
        \centering
        \includegraphics[width=0.99\columnwidth]{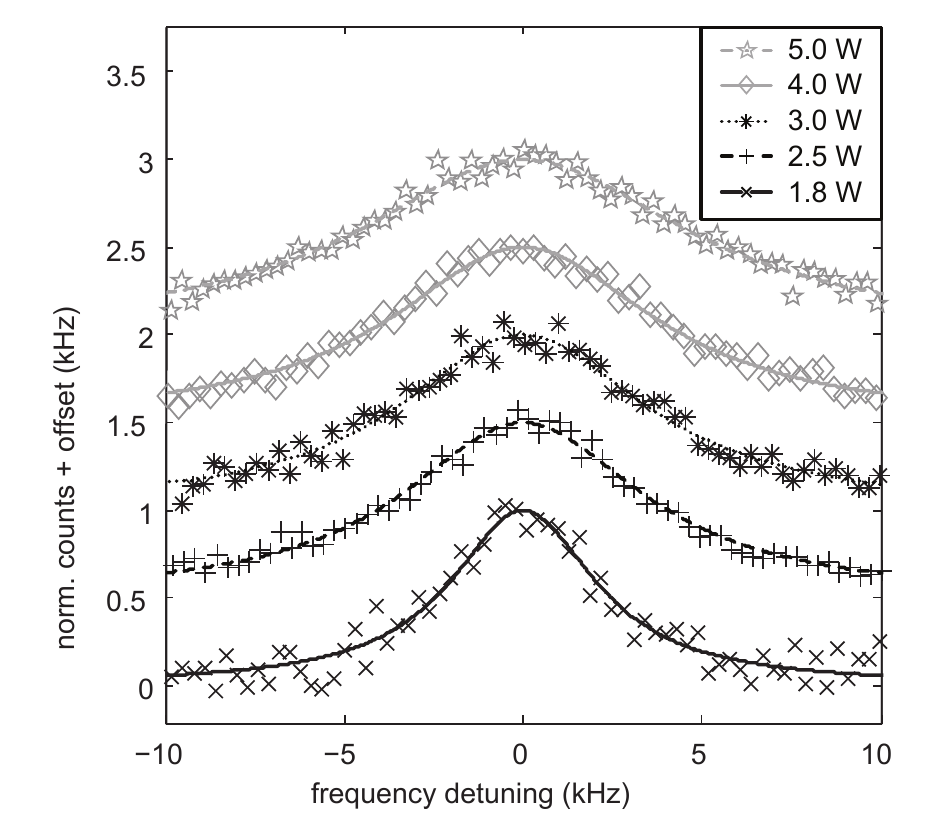}
        \caption[fig12]{RF-optical double resonance spectra. The RF power is varied while the
                    935~nm laser power remains constant near 5~\textmu W.
                    The transition width is reduced down to 5~kHz by diminishing the RF power.
                    A small width is necessary to detect small gradients. For a better visualization
                    offsets are added to the displayed spectra. The spectra are normalized and fitted with
                    Lorentzian profiles. }
        \label{rfspec}
    \end{figure}

    \begin{figure}
        \centering
        \includegraphics[width=0.99\columnwidth]{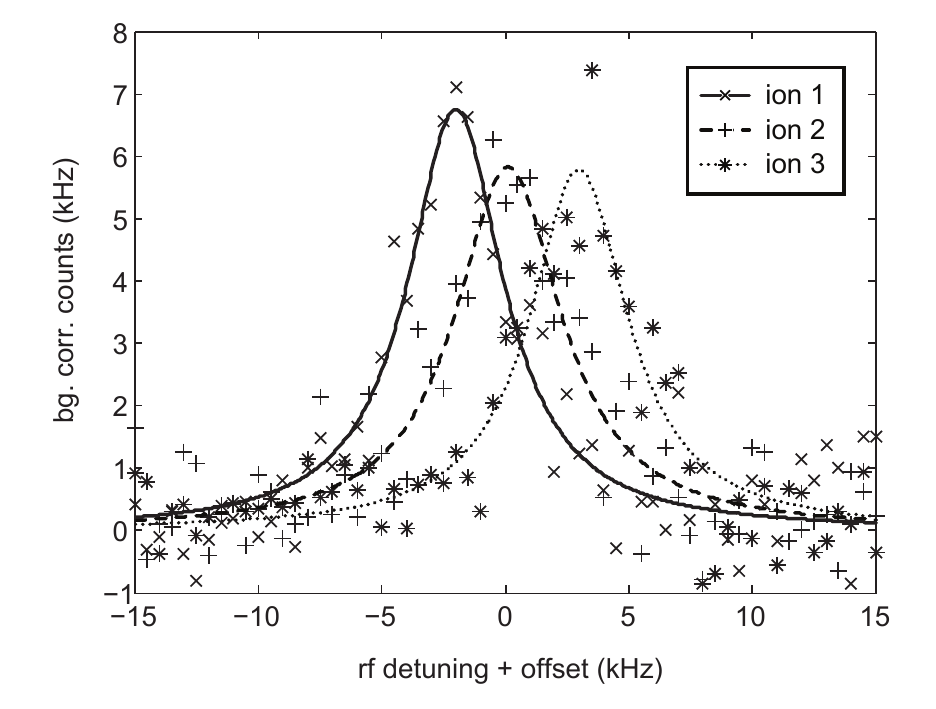}
        \caption[fig13]{ RF-optical double resonance spectra of a three-ion chain.
                    The measured gradient of 25~mT/m is composed of the gradient generated by a current
                    loop pair (14~mT/m) and the one due to the inhomogeneous, external magnetic field (11~mT/m).
                    Lorentzian functions fit the measured, background corrected data points.}
        \label{rfions}
    \end{figure}

\begin{figure}
        \centering
        \includegraphics[width=0.99\columnwidth]{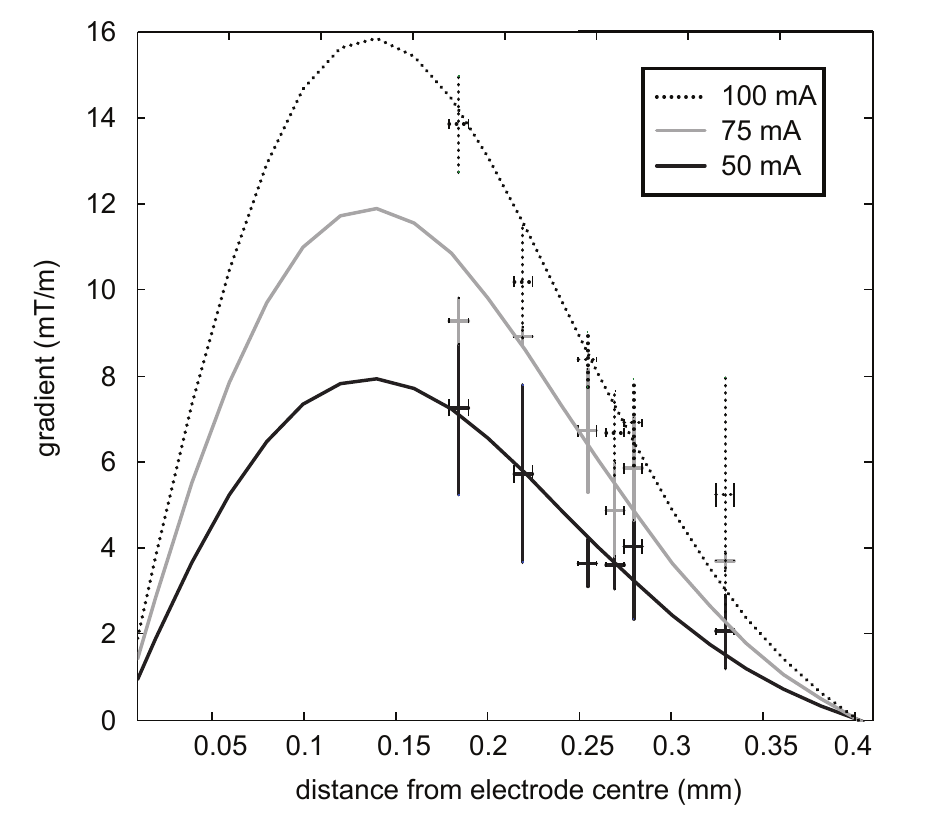}
        \caption[fig14]{Measured magnetic field gradients with currents
                    at 50~mA, 75~mA and 100~mA as a function of ion position. The lines
                    indicate the results of numerical simulations with no free parameters
                    taking into account the additional inhomogeneity from the permanent
                    magnet mounted outside the vacuum chamber.
                    The abscissa shows the distance from the centre of the current carrying
                    electrode pair. The error bars are statistical errors from a set of up
                    to 30 single measurements.}
        \label{gradcurr}
    \end{figure}

The quantization axis is defined by the field of a neodymium
permanent magnet positioned at a distance around 150~mm from the
trap center, which leads to a magnetic offset field near 0.6~mT.
RF-optical double resonance spectra, as shown in
fig.~\ref{rfspec}-\ref{gradcurr} are taken using the following
procedure: the repumper is split, manipulated with a half wave plate
and recombined using polarizing beam splitters. A half wave and a
quarter wave plate after the recombining polarizing beam splitter
(PBS) are adjusted such, that optical pumping is achieved using
light only from the first arm with an elliptical polarization as
described in sec.~\ref{sec:QubitInitializationAndReadout}.
Additional light from the second arm, which counteracts optical
pumping is chopped with a chopper-wheel to switch between cooling
cycles (no optical pumping) and spectroscopy cycles. During
spectroscopy cycles, the cooling laser and the repumper (leading to
optical pumping), RF manipulation and detection by the camera are
simultaneously active, leading to incoherent RF-optical double
resonance spectra with Lorentzian line shapes.

The amplitude and the width of the spectral lines are determined by
the power of the repumper laser and the RF radiation.
Fig.~\ref{rfspec} shows amplitude normalized spectra obtained by
scanning the RF frequency with constant laser power near 5~\textmu W
and varying RF power. For an RF power around 1.8~W we find the width
of the resonance to be around 5~kHz (FWHM). Under such conditions,
the narrow width of the resonance allows for a sensitive detection
of changes in the magnetic field strength (by changing the current)
or its inhomogeneity. The resonance frequency of the
$\sigma$-transition is given by eq.~\ref{eq:transition} and can thus
be changed by a variation of $B$ at the position of the ion. This
was checked by sending a current through the dc electrodes and
mapping out the resonance frequency as a function of the applied
current.

The magnetic field to which a given ion is exposed depends on its
position, and for a three-ion chain a small gradient will result in
three resonances whose center frequencies are shifted with respect
to each other. To demonstrate this, we send a current of up to
100~mA through two coils of the same segment and observe a splitting
as shown in fig.~\ref{rfions}. From this splitting and a measurement
of the ion separation, the gradient can be determined. The ion
separation is either directly determined from EMCCD images and the
calibrated magnification of the imaging system, or by measuring the
axial trap frequency and using the approximation for harmonic
potentials

\begin{equation}
\Delta y_2= \sqrt[3]{\frac{q^2}{2 \pi \epsilon_0 m \omega_{\rm
cm}^2}} \qquad \Delta y_3= \sqrt[3]{\frac{5}{8}} \Delta y_2
\end{equation}

($\Delta$ $y_2$ being the separation between two ions in a two-ion
chain and $\Delta$ $y_3$ being the separation between two ions in a
three-ion chain with centre-of-mass frequency $\omega_{\rm cm}$).
Numerical simulations with no free parameters and currents of 50~mA,
75~mA and 100~mA yield an expected maximum gradient of 16~mT/m and
agree well with the measurements carried out as discussed here for
various currents and positions of the ion string (see
fig.~\ref{gradcurr}).

    \subsection{RF addressing} \label{sec:RfAddressing}

    \begin{figure}
        \centering
        \includegraphics[width=0.99\columnwidth]{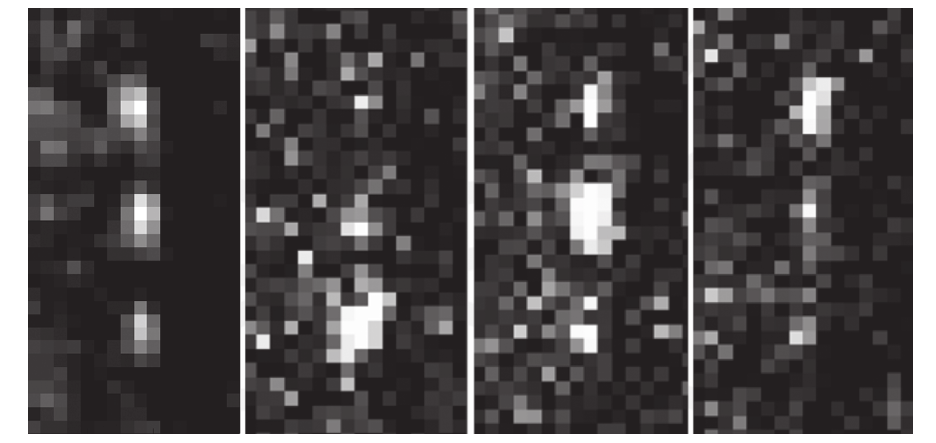}
        \caption[addressing]{Addressing an ion chain of three ions in frequency
                    space by tuning the applied RF frequency. The achieved separation in frequency
                    space is near 2.6~kHz with a current amount of 100~mA through an electrode
                    pair adding to an offset gradient generated by an external neodymium magnet.}
        \label{addressing}
    \end{figure}

With counter propagating currents of 100~mA through a current loop pair
and the additional inhomogeneity from the permanent magnet a
splitting of about 2.6~kHz is achieved near the maximal gradient
position of 130~\textmu m from the segment electrode centre for
axial trapping frequencies near 200~kHz in a three-ion chain. This
splitting is sufficient to individually address ions as is
demonstrated here for three ions in fig.~\ref{addressing}. Using the
average peak width of approximately 5~kHz from the measured data
presented in fig.~\ref{rfions}, we calculate the maximum undesired
excitation probability $p$ of a neighboring ion as 0.48~$\pm$~0.01
corresponding to a addressing fidelity of $F=1- p_{\rm{max}} =$
0.52~$\pm$~0.01 where $p_{\rm{max}}$ is the maximum unintended
excitation probability of all other ions. The corresponding fidelity
for a chain of two ions, under similar conditions would be
0.59~$\pm$~0.01, owing to the larger ion separation. This value can
be increased by applying a larger gradient, using lower trapping
frequencies, or by narrowing down the width of the resonance by
applying less RF and laser power. The constraints on the laser power
can be lifted entirely, when carrying out the RF-optical double
resonance spectroscopy in a coherent manner, such that the laser is
switched off during RF manipulation.

\section{Conclusion and outlook} \label{sec:Outlook}

We have discussed the design and production details for a segmented
planar trap with current loops that allow for tailoring of the axial
magnetic field gradient as well as the trapping potential and
demonstrated trapping of ytterbium ions in such a trap. The gradient
was demonstrated by applying RF-optical double resonance
spectroscopy, and a first demonstration of RF addressing in a planar
trap was shown. The size of the gradient was sufficient to carry out
proof-of-principle experiments. For quantum information processing,
however, a larger gradient will be necessary, as was demonstrated in
\cite{Khromova2012,Piltz2013,Wang2009}. An obvious solution will be
to apply higher currents. When currents are increased, the ultimate
limit for the current is the damage threshold of the current loops
(with the present trap this regime has not yet been explored).

When increasing the current sent through the loop structures, heat
management and thus cooling of the chip carrier becomes increasingly
important. The dissipated heat from ohmic losses in the carrier
limit at present possible currents to 100~mA, when restricted to
0.25~W power applied to the structure. The power limit of the
present setup was determined by analyzing the carrier temperature up
to approximately 2~W of applied heating power, allowing for
gradients near 60 mT/m with constant currents.

Pulsed currents, applied for short times, can create much larger
peak magnetic field gradients compared to gradients generated by
continuous currents while keeping the average power deposition low.
Possible peak gradients on the order of 1~T/m can be expected with
this trap setup.

As shown in the initial considerations of this paper,  higher
gradients for a fixed current can be reached by reducing the trap
dimensions, and from simulations we expect a gradient of up to
4~T/Am. The resulting gradient can be boosted up to 40~T/m when
applying pulsed currents of several Amperes with reduced trap
dimensions, which is on the order of present magnetic field
gradients in macroscopic traps using external magnets
\cite{Johanning2009b,Khromova2012}. With such large gradients, not
only addressing with high fidelity can be achieved, but also the
direct measurement of an effective spin-spin coupling, and its
application in quantum information science becomes possible
\cite{Khromova2012}.

\section*{Acknowledgments} \label{sec:Acknowledgements}

We would like to acknowledge M.~Epping for trap simulations,
M.~B\"ohm, D.~Sch\"afer-Stephani, K.~Watty, A.~Bablich,
P.~Haring-Bolivar, H.~Sch\"afer, E.~Ilichev, B.~Ivanov, and
S.~Zarazenkov for their support during chip production,
D.~Gebauer, D.~Junge, and A.~H.~Walenta for the production of the
chip carrier, our electrical and mechanical work shops, and
especially S.~Spitzer for his support regarding all electronics, and
T.~Collath, T.~F.~Gloger, D.~Kaufmann and P.~Kaufmann for providing
the laser system used in this work.

We acknowledge funding by the Bundesministerium f\"ur Bildung und
Forschung (FK 01BQ1012), and from the European Community's Seventh
Framework Programme (FP7/2007-2013) under Grant Agreement No. 270843
(iQIT) and No. 249958 (PICC).

$^{\dagger}$Present address: Max-Born-Institut f\"ur Nichtlineare
Optik und Kurzzeitspektroskopie, 12489 Berlin, Germany

\end{document}